\documentclass[longauth,letter,traditabstract]{aa}  
\usepackage{graphicx}
\usepackage{longtable}
\usepackage{multirow}
\usepackage{rotating}
\usepackage{natbib}
\usepackage{amsmath}
\usepackage{txfonts}

\newcommand{\um}{$\mu$m}                                 
\newcommand{\msun}{M$_{\odot}$}

\newcommand{\lsim}{\;\lower.6ex\hbox{$\sim$}\kern-7.75pt\raise.65ex\hbox{$<$}\;}
\newcommand{\gsim}{\;\lower.6ex\hbox{$\sim$}\kern-7.75pt\raise.65ex\hbox{$>$}\;}
\newcommand{\gl}{\;\lower.6ex\hbox{$<$}\kern-7.75pt\raise.65ex\hbox {$>$}\;}
\newcommand{\asec}{$^{\prime \prime}$}
\newcommand{\adeg}{$^{\circ}$}

\newcommand{\dder}{$\partial ^2$}
\newcommand{\higal}{\textbf{\textsl{Hi-GAL}}}
\begin{document}
\title{Clouds, filaments and protostars: the \textit{Herschel\thanks{Herschel is an ESA space observatory with science instruments provided by European-led Principal Investigator consortia and with important participation from NASA.}} \higal\ Milky Way}


\author{S. Molinari\inst{\ref{inst1}}\and
B. Swinyard\inst{\ref{inst2}}\and
J. Bally\inst{\ref{inst3}}\and
M. Barlow\inst{\ref{inst4}}\and
J.-P. Bernard\inst{\ref{inst5}}\and
P. Martin\inst{\ref{inst6}}\and
T. Moore\inst{\ref{inst7}}\and
A. Noriega-Crespo\inst{\ref{inst8}}\and
R. Plume\inst{\ref{inst9}}\and
L. Testi\inst{\ref{inst10},\ref{inst11}}\and
A. Zavagno\inst{\ref{inst12}}\and
A. Abergel\inst{\ref{inst13}}\and
Babar Ali\inst{\ref{inst14}}\and
L. Anderson\inst{\ref{inst12}}\and
P. Andr\'e\inst{\ref{inst15}}\and
J.-P. Baluteau\inst{\ref{inst12}}\and
C. Battersby\inst{\ref{inst3}}\and
M.T. Beltr\'an\inst{\ref{inst10}}\and
M. Benedettini\inst{\ref{inst1}}\and
N. Billot\inst{\ref{inst14}}\and
J. Blommaert\inst{\ref{inst16}}\and
S. Bontemps\inst{\ref{inst15},\ref{inst17}}\and
F. Boulanger\inst{\ref{inst13}}\and
J. Brand\inst{\ref{inst18}}\and
C. Brunt\inst{\ref{inst19}}\and
M. Burton\inst{\ref{inst20}}\and
L. Calzoletti\inst{\ref{inst56}}\and
S. Carey\inst{\ref{inst8}}\and
P. Caselli\inst{\ref{inst22}}\and
R. Cesaroni\inst{\ref{inst10}}\and
J. Cernicharo\inst{\ref{inst23}}\and
S. Chakrabarti\inst{\ref{inst58}}\and
A. Chrysostomou\inst{\ref{inst25}}\and
M. Cohen\inst{\ref{inst26}}\and
M. Compiegne\inst{\ref{inst27}}\and
P. de~Bernardis\inst{\ref{inst28}}\and
G. de~Gasperis\inst{\ref{inst29}}\and
A. M. di Giorgio\inst{\ref{inst1}}\and
D. Elia\inst{\ref{inst1}}\and
F. Faustini\inst{\ref{inst56}}\and
N. Flagey\inst{\ref{inst8}}\and
Y. Fukui\inst{\ref{inst32}}\and
G.A. Fuller\inst{\ref{inst33}}\and
K. Ganga\inst{\ref{inst34}}\and
P. Garcia-Lario\inst{\ref{inst35}}\and
J. Glenn\inst{\ref{inst3}}\and
P. F. Goldsmith\inst{\ref{inst37}}\and
M. Griffin\inst{\ref{inst38}}\and
M. Hoare\inst{\ref{inst22}}\and
M. Huang\inst{\ref{inst39}}\and
D. Ikhenaode\inst{\ref{inst54}}\and
C. Joblin\inst{\ref{inst5}}\and
G. Joncas\inst{\ref{inst41}}\and
M. Juvela\inst{\ref{inst42}}\and
J. M. Kirk\inst{\ref{inst38}}\and
G. Lagache\inst{\ref{inst13}}\and
J. Z. Li\inst{\ref{inst39}}\and
T. L. Lim\inst{\ref{inst2}}\and
S. D. Lord\inst{\ref{inst14}}\and
M. Marengo\inst{\ref{inst24}}\and
D. J. Marshall\inst{\ref{inst5}}\and
S. Masi\inst{\ref{inst28}}\and
F. Massi\inst{\ref{inst10}}\and
M. Matsuura\inst{\ref{inst4}, \ref{inst60}}\and
V. Minier\inst{\ref{inst15}}\and
M.-A. Miville-Desch\^enes\inst{\ref{inst13}}\and
L. A. Montier\inst{\ref{inst5}}\and
L. Morgan\inst{\ref{inst7}}\and
F. Motte\inst{\ref{inst15}}\and
J. C. Mottram\inst{\ref{inst19}}\and
T. G. M\"uller\inst{\ref{inst44}}\and
P. Natoli\inst{\ref{inst29}}\and
J. Neves\inst{\ref{inst43}}\and
L. Olmi\inst{\ref{inst10}}\and
R. Paladini\inst{\ref{inst8}}\and
D. Paradis\inst{\ref{inst8}}\and
H. Parsons\inst{\ref{inst43}}\and
N. Peretto\inst{\ref{inst33},\ref{inst15}}\and
M. Pestalozzi\inst{\ref{inst1}}\and
S. Pezzuto\inst{\ref{inst1}}\and
F. Piacentini\inst{\ref{inst28}}\and
L. Piazzo\inst{\ref{inst54}}\and
D. Polychroni\inst{\ref{inst1}}\and
M. Pomar\`es\inst{\ref{inst12}}\and
C. C. Popescu\inst{\ref{inst45}}\and
W. T. Reach\inst{\ref{inst8}}\and
I. Ristorcelli\inst{\ref{inst5}}\and
J.-F. Robitaille\inst{\ref{inst41}}\and
T. Robitaille\inst{\ref{inst24}}\and
J. A. Rod\'on\inst{\ref{inst12}}\and
A. Roy\inst{\ref{inst6}}\and
P. Royer\inst{\ref{inst16}}\and
D. Russeil\inst{\ref{inst12}}\and
P. Saraceno\inst{\ref{inst1}}\and
M. Sauvage\inst{\ref{inst15}}\and
P. Schilke\inst{\ref{inst30}}\and
E. Schisano\inst{\ref{inst1}, \ref{inst59}}\and
N. Schneider\inst{\ref{inst15}}\and
F. Schuller\inst{\ref{inst47}}\and
B. Schulz\inst{\ref{inst14}}\and
B. Sibthorpe\inst{\ref{inst38}}\and
H. A. Smith\inst{\ref{inst24}}\and
M. D. Smith\inst{\ref{inst49}}\and
L. Spinoglio\inst{\ref{inst1}}\and
D. Stamatellos\inst{\ref{inst38}}\and
F. Strafella\inst{\ref{inst21}}\and
G. S. Stringfellow\inst{\ref{inst3}}\and
E. Sturm\inst{\ref{inst44}}\and
R. Taylor\inst{\ref{inst50}}\and
M. A. Thompson\inst{\ref{inst43}}\and
A. Traficante\inst{\ref{inst29}}\and
R. J. Tuffs\inst{\ref{inst51}}\and
G. Umana\inst{\ref{inst52}}\and
L. Valenziano\inst{\ref{inst53}}\and
R. Vavrek\inst{\ref{inst35}}\and
M. Veneziani\inst{\ref{inst28}}\and
S. Viti\inst{\ref{inst4}}\and
C. Waelkens\inst{\ref{inst16}}\and
D. Ward-Thompson\inst{\ref{inst38}}\and
G. White\inst{\ref{inst2},\ref{inst57}}\and
L. A. Wilcock\inst{\ref{inst38}}\and
F. Wyrowski\inst{\ref{inst47}}\and
H. W. Yorke\inst{\ref{inst37}}\and
Q. Zhang\inst{\ref{inst24}}}

\institute{INAF-Istituto Fisica Spazio Interplanetario, Via Fosso del Cavaliere 100, I-00133 Roma, Italy
\email{sergio.molinari@ifsi-roma.inaf.it}\label{inst1}
\and
STFC, Rutherford Appleton Labs, Didcot, UK\label{inst2}
\and
Center for Astrophysics and Space Astronomy (CASA),
  Department of Astrophysical and Planetary Sciences, University of
  Colorado, Boulder, USA\label{inst3}
\and
Department of Physics and Astronomy, University College London, London, UK\label{inst4}
\and
Universit\'e de Toulouse , UPS, CESR, and CNRS, UMR5187, Toulouse, France \label{inst5}
\and
Department of Astronomy \& Astrophysics, University of Toronto, Toronto, Canada\label{inst6}
\and
Astrophysics Research Institute, Liverpool John Moores University, UK\label{inst7}
\and
Spitzer Science Center, California Institute of Technology, Pasadena, CA\label{inst8}
\and
Department of Physics \& Astronomy, University of Calgary, Canada\label{inst9}
\and
INAF - Osservatorio Astrofisico di Arcetri, Firenze, Italy\label{inst10}
\and
European Southern Observatory, Garching bei Muenchen, Germany\label{inst11}
\and
LAM, Universit\'{e} de Provence, Marseille, France\label{inst12}
\and
Institut d'Astrophysique Spatiale, Universit\'{e} Paris-Sud, Orsay, France\label{inst13}
\and
NASA Herschel Science Center, Caltech, Pasadena, CA\label{inst14}
\and
Laboratoire AIM, CEA/DSM - INSU/CNRS - Universit\'e Paris Diderot, IRFU/SAp CEA-Saclay, 91191 Gif-sur-Yvette, France\label{inst15}
\and
Institute for Astronomy, Katholieke Universiteit Leuven, Leuven, Belgium\label{inst16}
\and
LAB/CNRS, Universit\'e de Bordeaux, BP89, 33271 Floirac cedex, France\label{inst17}
\and
INAF - Istituto di Radioastronomia, Bologna, Italia\label{inst18}
\and
School of Physics, University of Exeter, Stocker Road, Exeter, EX4 4QL, UK\label{inst19}
\and
School of physics, University of New South Wales, Australia\label{inst20}
\and
Dipartimento di Fisica, Universit\'{a} del Salento, Lecce, Italy\label{inst21}
\and
School of Physics and Astronomy, University of Leeds, Leeds, UK\label{inst22}
\and
Centro de Astrobiologia, CSIC-INTA. Madrid, Spain\label{inst23}
\and
Harvard-Smithsonian Center for Astrophysics, Cambridge, MA\label{inst24}
\and
Joint Astronomy Center, Hilo, Hawaii\label{inst25}
\and
Radio Astronomy Lab., UCB, Berkeley, CA\label{inst26}
\and
Canadian Institute for Theoretical Astrophysics, University of Toronto,
  Toronto, Canada\label{inst27}
\and
Dipartimento di Fisica, Universit di Roma 1 "La Sapienza", Roma, Italy\label{inst28}
\and
Dipartimento di Fisica, Universit di Roma 2 "Tor Vergata", Roma, Italy\label{inst29}
\and
I. Physikalisches Institut der Universit\"at zu K\"oln, Z\"ulpicher Str. 77, 50937 K\"oln, Germany\label{inst30}
\and
Observat\'{o}rio Astronomico de Lisboa, Lisboa, Portugal\label{inst31}
\and
Department of Astrophysics, Nagoya University, Nagoya, Japan\label{inst32}
\and
Jodrell Bank Center for Astrophysics, School of Physics and Astronomy, University of
Manchester, Manchester, M13 9PL, UK\label{inst33}
\and
APC/Universit\'e Paris 7 Denis Diderot/CNRS, B\^atiment Condorcet, 10,
rue Alice Domon et L\'eonie Duquet, 75205 Paris Cedex 13, France\label{inst34}
\and
Herschel Science Centre, European Space Astronomy Centre, 
Villafranca del Castillo. Apartado de Correos 78, E-28080 Madrid, Spain\label{inst35}
\and
Jet Propulsion Laboratory, Pasadena, USA\label{inst37}
\and
School of Physics and Astronomy, Cardiff University, Cardiff, UK\label{inst38}
\and
National Astronomical Observatories, Chinese Academy of Sciences, Beijing, China\label{inst39}
\and
Departement de Physique, Universit\'e Laval, Qu\'ebec, Canada\label{inst41}
\and
Department of Physics, University of Helsinki, Finland\label{inst42}
\and
Centre for Astrophysics Research, Science and Technology Research Institute, University of Hertfordshire, Hatfield, UK\label{inst43}
\and
MPE-MPG, Garching bei M\"unchen, Germany\label{inst44}
\and
Jeremiah Horrocks Institute, University of Central Lancashire,
Preston PR1 2HE, UK\label{inst45}
\and
MPIfR-MPG, Bonn, Germany\label{inst47}
\and
Centre for Astrophysics \& Planetary Science, University of Kent, Canterbury, UK\label{inst49}
\and
Center for Radio Astronomy, University of Calgary, Calgary, Canada\label{inst50}
\and
Max-Planck-Institut f\"{u}r Kernphysik, Heidelberg, Germany\label{inst51}
\and
INAF-Osservatorio Astrofisico di Catania, Catania, Italy\label{inst52}
\and
INAF Istituto di Astrofica Spaziale e Fisica Cosmica, Bologna, Italy\label{inst53}
\and
Dipartimento di Scienza e Tecnica
dell'Informazione e della Comunicazione, Universit di Roma 1 "La Sapienza", Roma, Italy\label{inst54}
\and
ASI Science Data Center, I-00044 Frascati (Roma), Italy\label{inst56}
\and
Department of Physics and Astronomy, The Open University, Milton Keynes, UK\label{inst57}
\and
Astronomy Department, UCB Berkeley, CA\label{inst58}
\and
Dipartimento di Fisica, Universit\`{a} di Napoli "Federico II",  Napoli, Italy\label{inst59}
\and
Mullard Space Science Laboratory, University College London, Holmbury St. Mary,  Dorking, Surrey RH5 6NT, UK\label{inst60}
}

   \date{Received ; accepted}

 
  \abstract{We present the first results from the science demonstration phase for the \higal\ survey, the \textit{Herschel} key-project that will map the inner Galactic Plane of the Milky Way in 5 bands. We outline our data reduction strategy and present some science highlights on the two observed 2\adeg\ x 2\adeg\ tiles approximately centered at $l$=30\adeg\ and $l$=59\adeg. The two regions are extremely rich in intense and highly structured extended emission which shows a widespread organization in filaments. Source SEDs can be built for hundreds of objects in the two fields, and physical parameters can be extracted, for a good fraction of them where the distance could be estimated. The compact sources (which we will call Ôcores' in the following) are found for the most part to be associated with the filaments, and the relationship to the local beam-averaged column density of the filament itself shows that a core seems to appear when a threshold around $A_V\sim 1$ is exceeded for the regions in the $l$=59\adeg\ field; a $A_V$ value between 5 and 10 is found for the $l$=30\adeg\ field, likely due to the relatively higher distances of the sources. This outlines an exciting scenario where diffuse clouds first collapse into filaments, which later fragment to cores where the column density has reached a critical level. In spite of core L/M ratios being well in excess of a few for many sources, we find \textit{core} surface densities between 0.03 and 0.5 g cm$^{-2}$. Our results are in good agreement with recent MHD numerical simulations of filaments forming from large-scale converging flows.}


	\authorrunning{S. Molinari and the Hi-GAL Consortium}
	\titlerunning{The \higal\ Milky Way with \textit{Herschel}}
   \maketitle
%

\section{Introduction}

From the diffuse cirrus to the molecular clouds, onto the formation and death of stars, the Galactic Plane is the set where all the phases of the Galaxy life-cycle can be studied in context. Dust, best observed in the infrared and in the submillimeter, cycles through all these phases and is, as such, a privileged tracer for the Galactic ecology. IRAS \citep{neu84} and COBE \citep{mather90} were of tremendous importance in boosting the research in Galactic star formation and interstellar medium to the prominent positions they have today. As remote as they may now seem, however, these missions are only some 20 years away. Since then, a continuing explosion of Galactic Plane surveys, both in the mid-infrared at $\lambda \leq70$\um\ (\citealt{omo03}, \citealt{pri01}, \citealt{ben03}, \citealt{car09}) and in the submillimeter at $\lambda \geq 800$\um\ (\citealt{schu09}, \citealt{ros09}), are assembling a picture where the Galactic Plane has become accessible at sub-30\asec\ resolution over three decades of wavelength. The exception is the critical interval between 70 and 500\um\ where the bulk of the cold dust in the Galaxy emits and reaches the peak of its Spectral Energy Distribution (SED). The \higal\ Key-Project (Herschel infrared Galactic Plane survey) will fill this gap.

\higal\ is the key-project (KP) of the \textit{Herschel} satellite \citep{pilbratt10} that will use 343 hours observing time to carry out a 5-band photometric imaging survey at 70, 160, 250, 350, and 500\um\ of a $\mid b \mid \leq 1^{\circ}$-wide strip of the Milky Way Galactic Plane in the longitude range $-60^{\circ} \leq l \leq 60^{\circ}$. \higal\ is going to be the keystone in the multiwavelength Milky Way, opening up unprecedented opportunities with a promise of  breakthroughs in several fields of Galactic astronomy. A full description of the survey and its science goals are given elsewhere \citep{moli10a}. This contribution presents the first \higal\ data obtained in the \textit{Herschel} science demonstration phase (SDP) and describes a few of the main early results that will be detailed in other contributions in this volume.

\section{Observations and data reduction}

\begin{figure}[t]
\resizebox{\hsize}{!}{\includegraphics{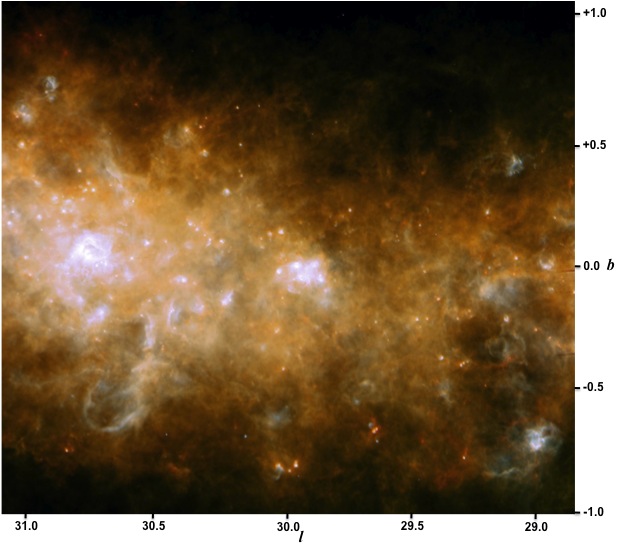}}
\caption{Three-color image (\textit{blue} 70\um, \textit{green} 160\um, \textit{red} 350\um) of the 2\adeg x2\adeg\ field around $l$=30\adeg}
\label{l30_color}
\end{figure}

\begin{figure}[t]
\resizebox{\hsize}{!}{\includegraphics{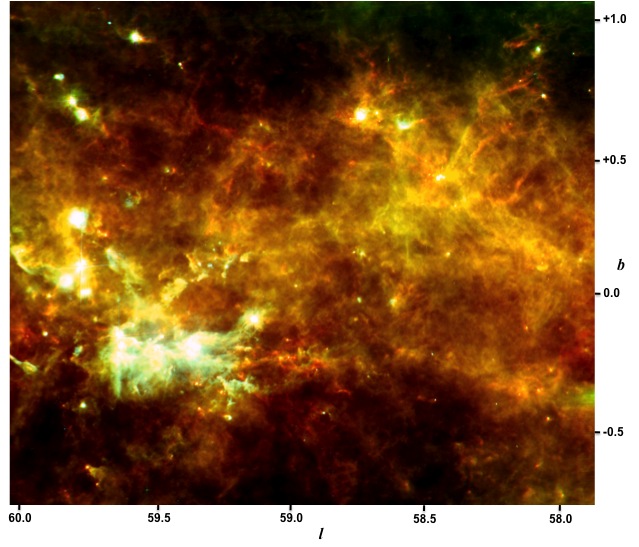}}
\caption{Three-color image (\textit{blue} 70\um, \textit{green} 160\um, \textit{red} 350\um) of the field around $l$=59\adeg}
\label{l59_color}
\end{figure}

The \textit{Herschel} PACS \citep{poglitsch10} and SPIRE \citep{griffin10} imaging cameras were used in parallel mode at 60\asec/s satellite scanning speed to obtain simultaneous 5-band coverage of two 2\adeg\ x 2\adeg\ fields approximately centered at [$l$, $b$]=[30\adeg, 0\adeg] and [59\adeg, 0\adeg]. The detailed description of the observation settings and scanning strategy adopted is given in \citet{moli10a}. Data reduction from archival data to Level 1 stage was carried out using the \textit{Herschel} Interactive Processing Environment (HIPE, \citealt{ott10}) using, however, custom reduction scripts that considerably departed from standard processing for PACS \citep{poglitsch10} and, to a lesser extent, for SPIRE \citep{griffin10}. Level 1 Time Ordered Data (TODs) were exported from HIPE into FITS files. Further processing including the map generation was carried out using dedicated IDL and FORTRAN codes. Saturation conditions were reached for all detectors only in SPIRE 250\um\ and 350\um\ images in correspondence with the 3 brightest peaks in the $l$=30\adeg\ field. The prescribed flux correction factors for PACS \citep{poglitsch10} and SPIRE \citep{swinyard10} were applied to the maps since their photometric calibration was carried out using the default calibration tree in HIPE. A detailed description of the entire data processing chain, including the presentation of the maps obtained in the five bands for the two observed fields, can be found in \citet{trafi10}. In the present letter we present in Figs. \ref{l30_color} and \ref{l59_color} the three-color images obtained using the 70, 160, and 350\um\ data ($l$=30\adeg\ and $l$=59\adeg, respectively). 

These amazing maps convey the immediate impression of extended filamentary structures dominating the emission on all spatial scales. Measurements of the standard deviation of the signal at all wavelengths in the lowest brightness regions of the $l$=59\adeg\ field yield average values a factor two higher than the sensitivity predictions for point source sensitivity from the HSpot time estimator for all bands except at  70\um\ where the predicted limit is effectively reached, confirming that the noise in our maps is dominated by the cirrus confusion at all wavelengths. A more detailed quantitative analysis is presented by \citet{martin10}.

\section{Results and science highlights}

\subsection{From IRDCs to mini-starburst and their impact on the ISM}

\textit{Herschel}'s ability to observe such large areas with unprecedented wavelength coverage and extraordinary signal dynamical range allows us to image simultaneously progenitors clouds for massive protoclusters to entire clusters of Young Stellar Objects (YSOs) in acitve star forming regions, while also measuring the effect of their strong stellar winds and powerful outflows on the surrounding medium. 

Infrared Dark Clouds (IRDCs) have received considerable attention in recent years (e.g. \citealt{rat06} and \citealt{peretto09}) as potential sites for precursors of cluster forming sites. Found in silhouette against the bright mid-IR background, they shine in emission with \textit{Herschel}. \citet{peretto10} shows how temperature effectively decreases from ambient values (20-30K) down to T=8-15K toward the densest ($\sim 10^{23}$cm$^{-2}$) peaks of these objects, resolving further temperature substructures that can be proxies for subsequent fragmentation.

At the other end of the massive star formation timeline, we find W43, visible in the left portion of Fig. \ref{l30_color},  as an outstanding case of Galactic mini-starburst. Detailed SED construction and luminosity estimates allow us to assess the very early evolutionary stage of the most luminous and massive YSOs in the region. It is remarkable how the same images show a prominent ridge extending southward which encompasses a 70pc-wide large cavity excavated by the W43 cluster and which possibly triggers further star formation \citep{bally10}. Triggered star formation in less extreme environments can also be studied in statistically significant fashion modeling the SED of the sources found in correspondence of the multitude of H{\sc ii}-driven bubble-like structures found in the images, as shown for the bubble N49 by \citet{zavagno10}.

Feedbacks from massive star formation, together with the intricate relationship between the interstellar radiation field and molecular clouds, are at the origin of the observed complexity of the ISM emission structure, where temperature ranges from $\sim$10K of pre-stellar cores to the $\sim$40-50K of the photodissociation regions \citep{bernard10}.

\subsection{Census of compact sources}
\label{census}

The extraction of compact sources is quite a challenging task in these fields, which we faced using a novel approach  based on the study of the multidirectional second derivatives in the image to aid in source detection and size estimate, and subsequent constrained multi-Gaussian fitting. This approach greatly increases the dynamical range between compact sources and diffuse emission, irrespective of the local absolute value of the emission. The method is fully described elsewhere \citep{moli10b} and has been applied for this first attempt to generate source catalogs. As the thresholding for source detection is done on the \textit{curvature} image \citep{moli10b}, the S/N of the detected sources is determined \textit{a posteriori} measuring the ratio of the source peak flux over the $rms$ of the residuals after the Gaussian fit. Source catalogs were generated for the two fields and for the 5 bands and are made available in tabular form in the online version of the paper. Catalogs completeness was estimated with artificial source injection experiments, and the peak flux levels for 80\% completeness for the 70, 160, 250, 350 and 500\um\ photometry are [0.5, 4.1, 4.1, 3.2, 2.5] Jy/beam for the $l$=30\adeg\ field, and [0.06, 0.9, 0.7, 0.7, 0.8] Jy/beam for the $l$=59\adeg\ field. The difference is entirely compatible with the very different intensity regimes of the underlying diffuse emission in the two fields.

Estimating the source's physical properties requires that detection in the various band catalogs are merged in coherent SEDs, a process that can only be done coarsely in this early stage, but which is nonetheless useful for  isolating 528 sources in the $l$=30\adeg\ field and 444 sources in the $l$=59\adeg\ field (see \citealt{elia10}). The two observed fields encompass emission from regions at very different distances. In a considerable effort, which involved a critical re-evaluation of available data and evidence, and the collection of additional data for hundreds of previously unknown objects, \citet{russeil10} provide recommended distances for a fraction of the detected sources (312 out of 528,  and 91 out of 444 sources for the two fields, respectively) for which the derivation of masses and luminosities is possible. Adopting standard prescriptions for Class 0  classification ($L_{\lambda \geq 350\mu m}/L_{bol}\geq 0.005$, \citealt{andre00}) results in almost the totality of sources being Class 0 (90 out of 91 sources in $l$=59\adeg\ and 306 out of 312 in $l$=30\adeg, see \citealt{elia10}).

\subsection{Filamentary star formation}

\begin{figure}[t]
\resizebox{\hsize}{!}{\includegraphics{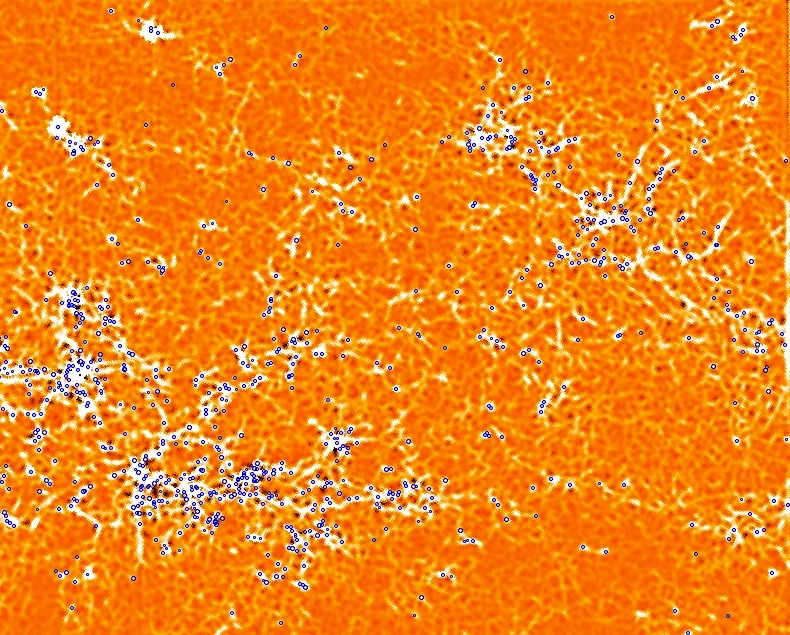}}
\caption{Multidirectional second-derivative image (see text) of the $l$=59\adeg\ field at 250\um. The blue circles represent the compact sources detected at 250\um. The filamentary structure of the ISM appears at various levels of intensity, i.e. curvature, and it is striking how the detected compact objects are for the most part distributed only along the brightest filaments.}
\label{l59_filaments}
\end{figure}

The most extraordinary feature exhibited by the \textit{Herschel} maps is the \textit{ubiquitous} pattern of filaments in the ISM structure. This is more apparent when we enhance the contrast of the filaments using the same method (see \S\ref{census}) as used for the source detection \citep{moli10b}. Here we start from the \dder\ derivatives carried out in four directions (x, y, and the two diagonals), as for the standard detection method, and then create another image F of the same size so that the maximum curvature is selected for each pixel: $F_{ij} = max [ \partial ^2 {\rm{x}} _{ij}, \partial ^2 {\rm{y}} _{ij}, \partial ^2 {\rm{D_{1}}} _{ij},  \partial ^2 {\rm{D_{2}}} _{ij} ]$. In this way we are following the direction of maximum curvature pixel-by-pixel for all compact features in the image. We show in fig.\ref{l59_filaments} the result of this processing on the $l$=59\adeg\ field at 250\um. 

\begin{figure}[t]
\resizebox{\hsize}{!}{\includegraphics{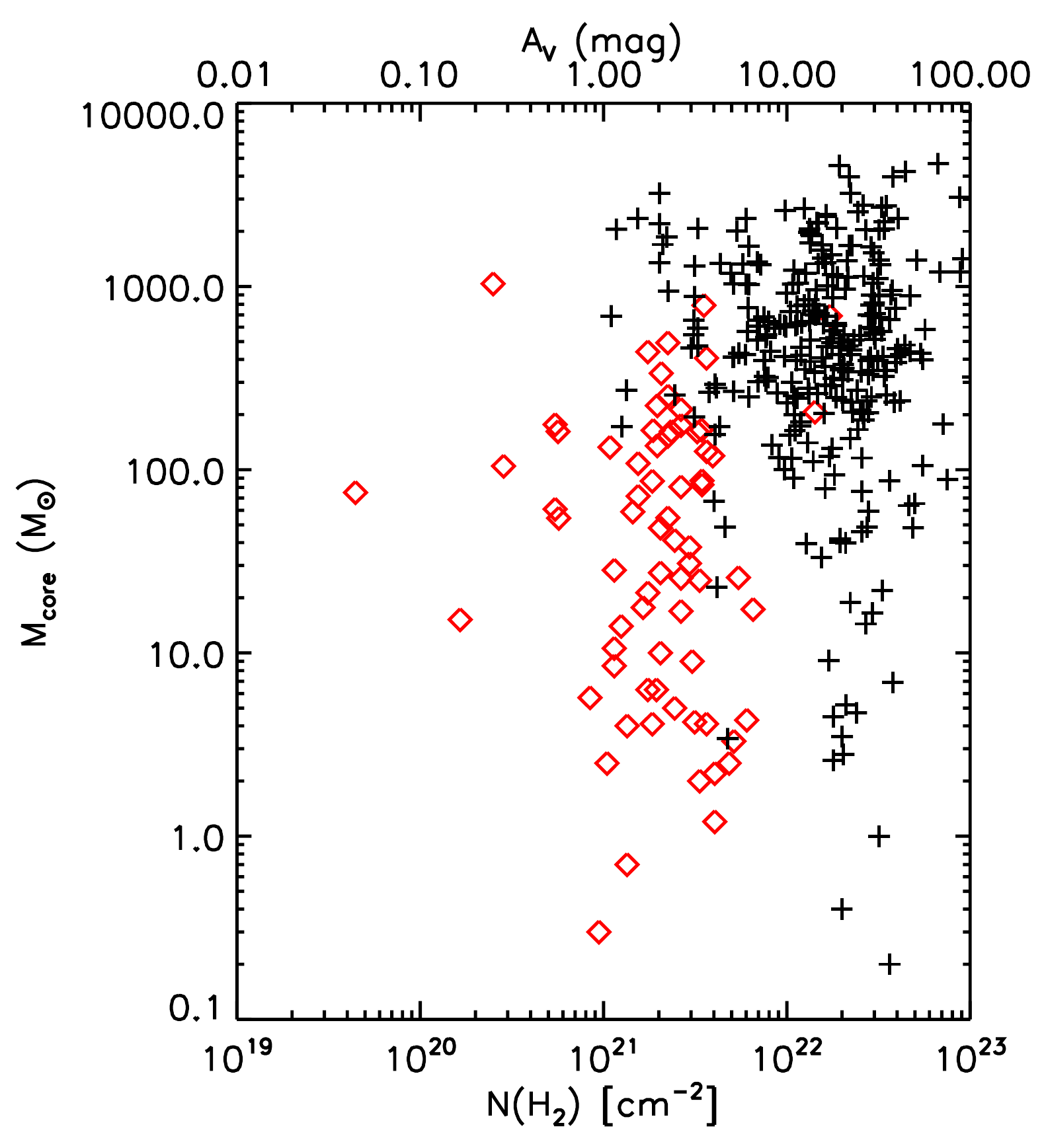}}
\caption{Mass of detected cores as a function of the hosting filament local column density, for the cores where the distance is known and the mass could be estimated. The red diamonds represent the cores in the $l$=59\adeg\ region, while the plus signs represent the cores from the $l$=30\adeg\ region. The top X-axis represents the equivalent A$_V$.}
\label{mcore-mback}
\end{figure}

The maps clearly show an interconnected maze of filaments at different levels of brightness (e.g. different levels of emission intensity and curvature), and the striking aspect is that the compact sources detected at 250\um\ are distributed for the most part along the brightest filaments. Interestingly, a similar scenario was also reported for Taurus by \citet{gold08} where the physical conditions and spatial scales involved are radically different. Since the source integrated fluxes are estimated by fitting Gaussians on top of planar plateaus, the values of the local background at every wavelength are a by-product of our source extraction and, after applying the absolute correction factors as recommended by \citet{bernard10} and subtracting the foreground contribution estimated using \citet{bohlin75},  can be used to estimate the local beam-averaged column density in the hosting filaments. The relationship between the mass of the detected cores (when the SED is reliable and the distance is known, see previous paragraph) and the local beam-averaged H$_2$ filament column density is reported in Fig. \ref{mcore-mback}. The points for the $l$=59\adeg\ field mostly lie in the range of column densities ($10^{21}cm^{-2}\leq\ N(H_2) \leq 10^{22} cm^{-2}$) that corresponds to $1mag\leq A_V \leq 10mag$, values that are entirely reasonable for the transition regime between diffuse ISM and dense molecular clouds (\citealt{cambresy99}, \citealt{snow06}). Higher values of $N(H_2)$ are found for the points for the $l$=30\adeg\ field, most likely due to the larger relative distances of the sources in this latter field. The core masses are spread between 1 and 10$^4$ \msun, with no indication of a correlation between the two quantities. The strong impression, however, is that of a threshold at $A_V\sim\ 1mag$ for the $l$=59\adeg\ field above which dense cores are found, a threshold that is evidently exceeded only in bright filaments (Fig. \ref{l59_filaments}). More in particular, the $A_V\sim\ 1mag$ threshold corresponds to $\sim$ 17\msun\ pc$^{-2}$ in molecular hydrogen, which is suprisingly close to the 10\msun\ pc$^{-2}$ value that \citet{krum09} find critical for the dust content in HI clouds to efficiently shield the cloud interior from external FUV field and allow effective H$_2$ formation. It is tempting to relate the appearance of clumps to an extinction regime where the H{\sc i}/H$_2$ boundary shields the cloud interiors from interstellar FUV field, causing the photoelectric heating efficiency to drop considerably and causing in turn a drop in dust and gas temperature \citep{th85}. This threshold value seems to be of the order of $A_V\sim 5\div 10$ for the $l$=30\adeg\ field most likely due to the relatively larger distances of sources in this field \citep{russeil10}.


The ubiquitousness of dense filaments in the ISM, the high degree of association between bright filaments and cores, and the suggestion of a column density threshold for the appearance of dense cores, all appear to coherently support a formation scenario that starts with the condensation of diffuse clouds into long filaments. As the column density increases, a threshold is exceeded and denser star-forming (or potentially star forming) condensations start to appear. A preliminary association with Spitzer 24\um\ counterparts \citep{elia10} suggests that our detected sources may be a mixture of protostellar and pre-stellar objects, although more work will be needed to ascertain the composition of this mixture.

Since our source extraction also yields a measure of the core sizes, we are in the position to estimate their surface  density $\Sigma$. These values appear to be on average a factor 3-5 higher than the column densities of the underlying filaments as reported in Fig. \ref{mcore-mback},  spanning a range of 0.03 g cm$^{-2} \leq \Sigma  \leq 0.5$g cm$^{-2}$, after changing units, with a mean value of 0.1 g cm$^{-2}$. It is puzzling that the number of cores exceeding L/M ratios of a few (the exact number depending on the core masses), corresponding to the critical surface density threshold of 1 g cm$^{-2}$ for the formation of massive stars \citep{krum08}, is not consistent with very few of the cores actually exceeding that critical threshold (see also \citealt{elia10}). The difference of a factor $\sim$2 between the dust opacities that we used compared to \citet{krum08} is not sufficient to reconcile this apparent discrepancy. This result deserves more attention and needs to be confirmed in the future with more detailed and accurate analysis.

Testing of large-scale "dynamical" star formation scenarios (e.g., \citealt{har01}), where filaments are formed in the post-shock regions of large H{\sc i} converging flows, is one of the original science goals of \higal. It is remarkable how the predictions from recent MHD numerical simulations \citep{banerjee09} of formation and subsequent fragmentation of filaments, agree with our results. Besides the morphological resemblance of these simulations with the structures we see in our \textit{Herschel} maps (Fig. \ref{l59_filaments}), there is striking agreement of their predictions with the N(H) regime we measured for our core-hosting filaments, as well as with the mass regime of the cores being formed.

Instability and fragmentation of dense filaments has also been investigated in the context of helical magnetic fields enclosing the filaments by \citet{fiege00}; interestingly, the models predict the formation of regularly spaced condensations at  length scales that depend on the properties of the magnetic field, the velocity dispersion, and density of the filament. The predicted length scale for filament velocity dispersion of 0.5 km s$^{-1}$ and density of 10$^4$ cm$^{-3}$ is 2.8 pc, and curiously this is not at all far from the median distance of each source to its nearest neighbor: $\sim$1.8pc for the 250\um\ sources in the $l$=59\adeg\ field for the sources' average distance \citep{russeil10}. It is also interesting that the typical fragmentation length scale decreases with increasing filament density, in broad qualitative agreement with a higher spatial density of sources in the brightest filaments. This clearly deserves further investigation to be confirmed as a viable hypothesis.

\section{Conclusions}

The first science highlights presented in this paper, as well as in the accompanying papers in this volume and elsewhere, show that owing to its optimal use of unique \textit{Herschel} characteristics of wavelength coverage, spatial resolution and mapping speed, the \higal\ survey has the potential to lead to a quantum leap in our understanding of large-scale Galactic star formation from cloud to cluster-forming clump formation and of the evolution of protoclusters and  massive protostars.

\textit{The outstanding feature emerging from these first images is the impressive and ubiquitous ISM filamentary nature. Dense cores seem to appear when a certain beam-averaged column density threshold is exceeded in close spatial association with these filaments.}

\begin{acknowledgement}
Data processing and map production has been possible thanks to generous support from the Italian Space Agency via contract I/038/080/0. We are indebted to Nicola Giordano for the production of the trichromatic overlay images. Data presented in this paper were also analyzed using ÒThe Herschel Interactive Processing Environment (HIPE),Ó a joint development by the Herschel Science Ground Segment Consortium, consisting of ESA, the NASA Herschel Science Center, and the HIFI, PACS, and SPIRE consortia. 
\end{acknowledgement}


\end{document}